\documentclass[10pt,conference]{IEEEtran}
\usepackage{booktabs}
\usepackage{listings}
\usepackage{graphicx}
\usepackage{tabularx}

\usepackage{cite}
\usepackage{hyperref}

\ifCLASSINFOpdf
\else
\fi
\usepackage{array}
\usepackage{url}

\usepackage[utf8x]{inputenc}

\begin{document}
%
\title{FixMe: A GitHub Bot for Detecting and Monitoring On-Hold Self-Admitted Technical Debt}



%
\author{\IEEEauthorblockN{
Saranphon Phaithoon\IEEEauthorrefmark{1},
Supakarn Wongnil\IEEEauthorrefmark{1},
Patiphol Pussawong\IEEEauthorrefmark{1}, \\
Morakot Choetkiertikul\IEEEauthorrefmark{1},
Chaiyong Ragkhitwetsagul\IEEEauthorrefmark{1},
Thanwadee Sunetnanta\IEEEauthorrefmark{1}, \\
Rungroj Maipradit\IEEEauthorrefmark{2},
Hideaki Hata\IEEEauthorrefmark{3}, and
‪Kenichi Matsumoto\IEEEauthorrefmark{2}
}
\IEEEauthorblockA{\IEEEauthorrefmark{1}Faculty of Information and Communication Technology (ICT),
Mahidol University,
Nakhon Pathom, Thailand\\
\IEEEauthorrefmark{2}Nara~Institute~of~Science~and Technology (NAIST), Nara, Japan\\
\IEEEauthorrefmark{3}Shinshu University, Nagano, Japan
}
\IEEEauthorblockA{
Email: saranphon.pha@gmail.com, supakarn.won@hotmail.com, patiphol.pus@gmail.com\\
\{morakot.cho, chaiyong.rag,  thanwadee.sun\}@mahidol.ac.th \\
\{maipradit.rungroj.mm6, matumoto\}@is.naist.jp, 
hata@shinshu-u.ac.jp
}
}


\maketitle

\begin{abstract}
Self-Admitted Technical Debt (SATD) is a special form of technical debt in which developers intentionally record their hacks in the code by adding comments for attention. Here, we focus on issue-related ``On-hold SATD'', where developers suspend proper implementation due to issues reported inside or outside the project. When the referenced issues are resolved, the On-hold SATD also need to be addressed, but since monitoring these issue reports takes a lot of time and effort, developers may not be aware of the resolved issues and leave the On-hold SATD in the code.
In this paper, we propose FixMe, a GitHub bot that helps developers detecting and monitoring On-hold SATD in their repositories and notify them whenever the On-hold SATDs are ready to be fixed (i.e. the referenced issues are resolved). The bot can automatically detect On-hold SATD comments from source code using machine learning techniques and discover referenced issues. When the referenced issues are resolved, developers will be notified by FixMe bot. The evaluation conducted with 11 participants shows that our FixMe bot can support them in dealing with On-hold SATD. FixMe is available at \url{https://www.fixmebot.app/} and  FixMe's VDO is at \url{https://youtu.be/YSz9kFxN\_YQ}.
\end{abstract}


%
\IEEEpeerreviewmaketitle

\section{Introduction}

On-hold Self-admitted Technical Debt (On-hold SATD) implies the debt that is created when the implementation need to wait (i.e., be held) for a resolution of a referenced issues in source code comments. These dependencies can refer to same project's or other project's issues. Those On-hold SATDs are supposed to be fixed after their referenced issues are resolved, so called Ready-to-be-fixed On-hold SATD \cite{Maipradit2018}. However, the study shows that a number of Ready-to-be-fixed On-hold SATDs have actually never been fixed  \cite{Maipradit2018,Maipradit2020,LiSATDIssueTracker}.

\begin{figure}[h]
    \centering
    \includegraphics[width=\linewidth]{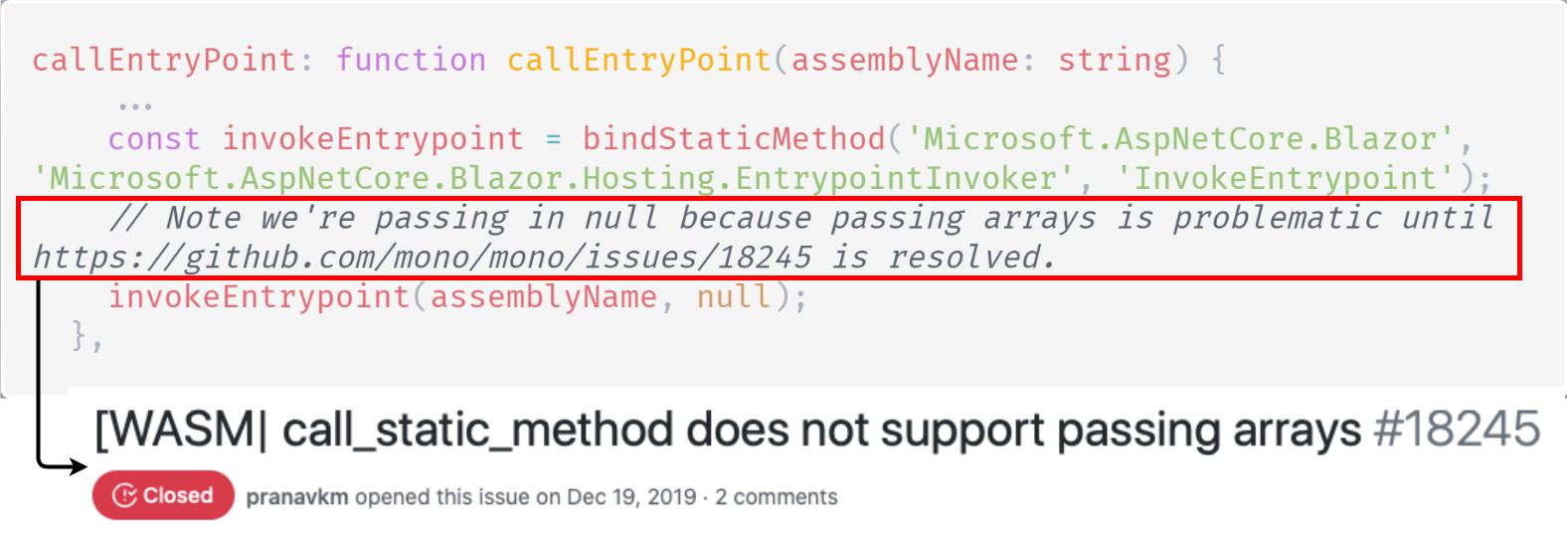}
    \caption{Example of Ready-to-be-fixed On-hold SATD in the Mono project}
    \label{fig:comment}
\end{figure}

A software module usually relies on numerous dependencies such as calling functions from within a project or across projects. When those issue referencing in source code comments have been resolved, developers are not aware of such issue resolving. Those On-hold SATDs thus do not usually be maintained properly. For example, the source code comments\footnote{\url{https://bit.ly/3p6hGuZ}} in the Mono project\footnote{\url{https://www.mono-project.com/}} shows that the problem of passing arrays can be fixed after the issue ID \texttt{18245}\footnote{\url{https://bit.ly/3v2FToi}} was resolved.
However, the issue was resolved but the code have not been fixed (see Figure \ref{fig:comment}). Therefore, this manual monitoring and maintaining of referenced issues are tedious and they require tremendous~ effort. 

In order to fill this gap and help code reviewers and developers overcome these challenges, we present ``FixMe bot'', a GitHub bot that help code reviewers and developers to detect and monitor On-hold SATDs from source code comments and inform them whenever an On-hold SATD is Ready-to-be-fixed by adopting the existing On-hold SATD detection approach from Maipradit et al. \cite{Maipradit2018}.

\section{Related work}
Recently, there is an existing On-hold SATD detection model from Maipradit et al. \cite{Maipradit2020} who applied regular expressions to detect issue references in source code comments and proposed a machine learning model to identify source code comments as either ``On-hold SATD'' (the referenced issue is used to indicate the need to wait for issue resolution) or ``Cross reference'' (the referenced issue is used to document the code) \cite{Maipradit2018}. 
Additionally, Huang et al.~used text mining to identify SATD in eight open source projects \cite{Huang2018}. 
Potdar and Shihab \cite{6976075} reported 62 common patterns that can be used to identify SATD.
Maldonado et al. \cite{7820211} present an approach to automatically identify design and requirement SATD using Natural Language Processing (NLP) on code comments.
Ren et al. \cite{10.1145/3324916} proposed a Convolutional Neural Network (CNN) to classify code comments into SATD or non-SATD. 
Yu et al. \cite{9226105} introduce a framework that reduce human effort in identifying SATD by using 2 steps. First, the framework uses patterns to identify easy-to-find SATDs. After that, it uses machine learning to assist human expert in manually identifying hard-to-find SATDs.
SonarQube \cite{sonarqube} is a platform integrated with external tools like GitHub that provides the capability to view and analyze technical debt in term of number of working days required to solve all technical debt. Just-In-Time Bot (JITBot) \cite{9286007} is a GitHub application that automatically provide feedback about riskiness of each commit and suggest risk mitigation plans to developers.
However, there is no On-hold SATD detection tool that is fully integrated into software development process such as a code review in practice.

 

\section{An overview of our FixMe bot}

\begin{figure}[htb!]
    \centering 
    \includegraphics[width=\linewidth]{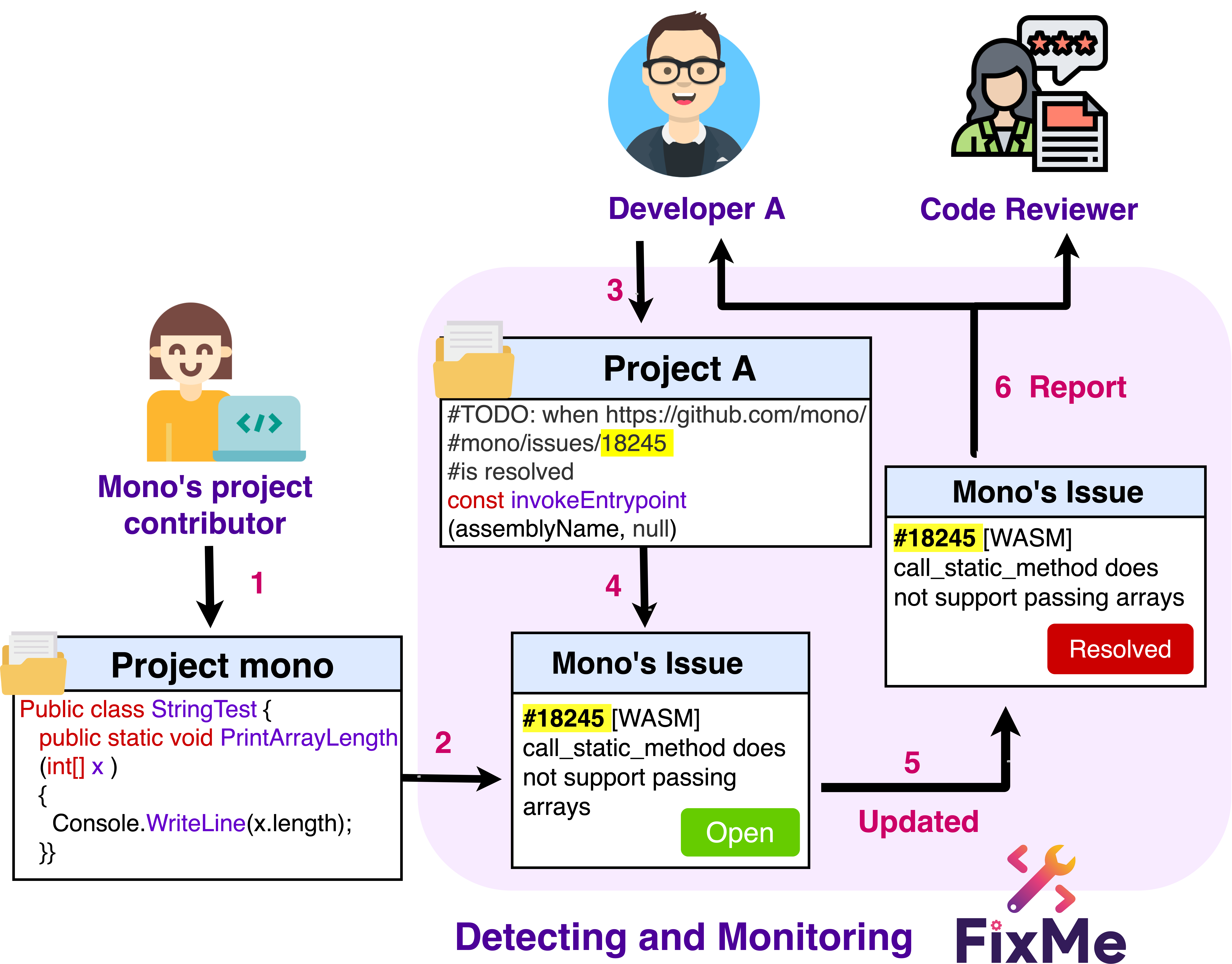}
    \caption{The scenario of using FixMe}
    \label{fig:proposed-scenario}
\end{figure}

\begin{figure}
    \centering
    \includegraphics[width=\linewidth]{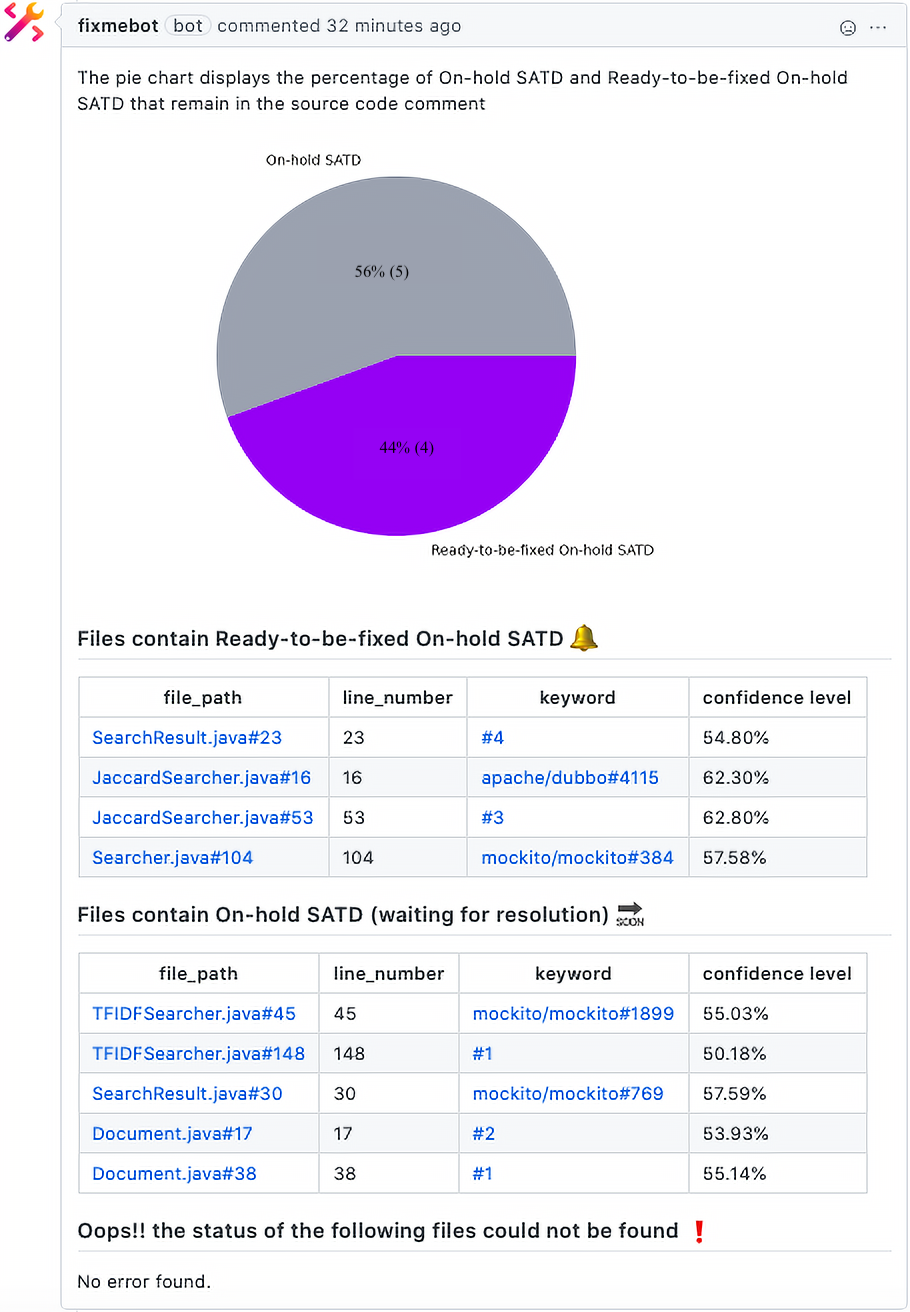}
    \caption{Example of pull request and commit comments posted by FixMe bot}
    \label{fig:pull_request}
\end{figure}

\begin{figure}
    \centering
    \includegraphics[width=\linewidth]{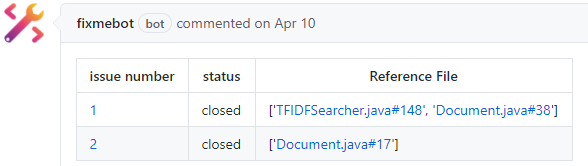}
    \caption{Example of issue comment posted by FixMe bot}
    \label{fig:issue_creation}
\end{figure}

We explain how FixMe bot works in Figure \ref{fig:proposed-scenario}. FixMe bot analyzes source code in GitHub repositories to find code comments. Using the existing On-hold SATD detection model~\cite{Maipradit2018}. FixMe bot can automatically determine whether a code comment is an On-hold SATD or an issue referencing comment (e.g., to note related issues in the code). GitHub issue reports that are mentioned in On-hold SATD comments are then discovered using regular expressions. After extracting a list of issue reports and On-hold SATD code comment locations, the bot monitors the issue status and notify software practitioners when those issue are resolved. There are three form of outputs that FixMe bot informs the developers of the Ready-to-be-fixed SATDs: 1) posting a comment containing a list of Ready-to-be-fixed On-hold SATDs on a pull request with the confidence level that shows the estimated probability obtained from the On-hold SATD classification model (Figure \ref{fig:pull_request}), 2) posting a comment on a commit that shows Ready-to-be-fixed On-hold SATDs found on the commit (i.e., code changes), and 3) creating a new issue report to list all the Ready-to-be-fixed On-hold SATDs in a project (Figure \ref{fig:issue_creation}). In addition, FixMe can be also configured via \url{https://www.fixmebot.app}. For example, the developers can select the output methods, select branches to monitor, and add new patterns of regular expressions.




\section{System Architecture}

\begin{figure*}
    \includegraphics[width=0.95\textwidth]{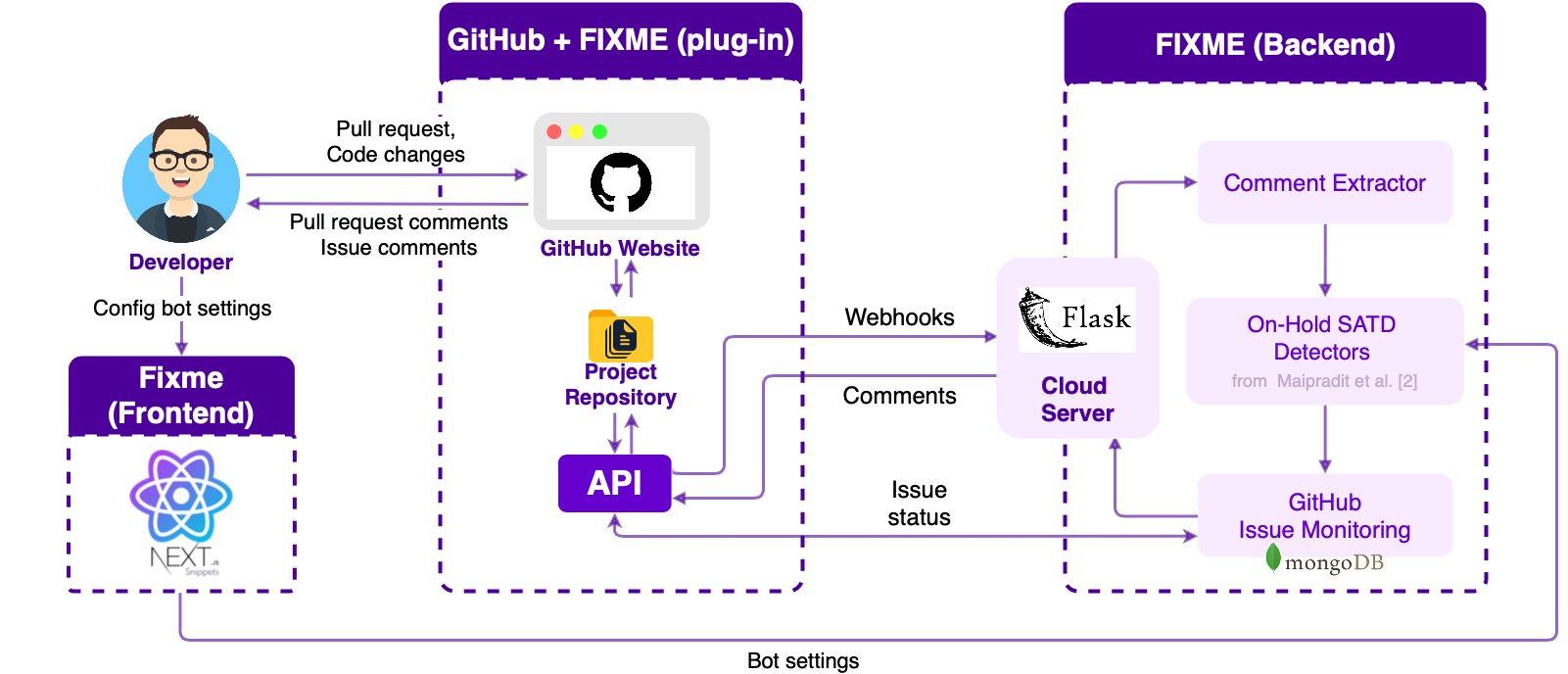}
    \caption{The overview architecture of FixMe}
    \label{fig:system-architecture}
\end{figure*}


Figure \ref{fig:system-architecture} shows the FixMe's architecture which consists of three components: FixMe GitHub plug-in, FixMe backend, and FixMe frontend (configuration page). First, a project owner (e.g., repository maintainer) needs to install the FixMe plug-in into their GitHub repository. Whenever project contributors push a commit or create a new pull request, GitHub sends a webhook to trigger the FixMe backend module. The FixMe bot then make a request to GitHub's API for posting outputs based on the user's configuration. 



\subsection{FixMe GitHub Plug-in}
GitHub provides the GitHub API to allow integrations from third party applications to perform their functionalities. The FixMe GitHub plug-in is responsible for creating webhooks from GitHub to establish the connection between GitHub and the FixMe bot's back-end~system.


\subsection{FixMe Backend}

The backend of FixMe bot contains three components: Comment extraction, On-hold SATD detection, and GitHub issue monitoring (Figure~\ref{fig:system-architecture}). 

\subsubsection{Comment Extraction}
The FixMe bot can extract source code comments from a specific commit in a repository when the bot receives a webhook from the FixMe plug-in. The bot then collects all files referenced in those requests and filters only source code files based on file extensions. The bot then uses the comment parser introduced in Maipradit et al. \cite{Maipradit2018} which technically uses regular expressions to extract source code comments through all source code files. Currently, FixMe supports only Java language.  
    
    
\subsubsection{On-hold SATD Detection}
The extracted code comments are then fed into the On-hold SATD detection model adopted from Maipradit et al. \cite{Maipradit2018} to determine whether the extracted code comments are On-hold SATDs or not. The model is trained from N-gram features of the code comments from 10 Java open source projects using Auto-sklearn version 0.9.0 and Python version 3.7.10. It shows that the Extra Tree algorithm was the best performer to classify On-hold SATDs. The model achieves over 0.8 F-measure and 0.9 AUC scores \cite{Maipradit2018}. 

\subsubsection{GitHub Issue Monitoring}
The referencing issues contained in those identified On-hold SATD comments are extracted and recorded into FixMe's database. We used a \emph{cron job}, a built-in Linux utility, to periodically run the issue monitoring script. Thus, whenever referenced issues are resolved, FixMe invokes the GitHub API to notify software practitioners.


\subsection{FixMe Configuration}

There are three parts that users can configure FixMe: selecting On-hold SATD detection techniques, specifying target branch(es), and choosing output posting methods. In On-hold detection methods, FixMe allows users to define their own regular expression patterns to detect On-hold SATD comments. Thus, they can modify the detection methods to match with their own development practices. For example, a team can define a common comment patterns that reflect the On-hold SATD comment (e.g., \verb!(after|once)\s+(issue \d+) is resolved!).
FixMe also supports multiple-branch monitoring and the users can select a number of target branches to be monitored by FixMe. Lastly, the users can select three output methods: pull request comment, commit comment, and issue creation.

\section{Evaluation and Result}

    \begin{figure}
        \centering
        \includegraphics[width=\linewidth]{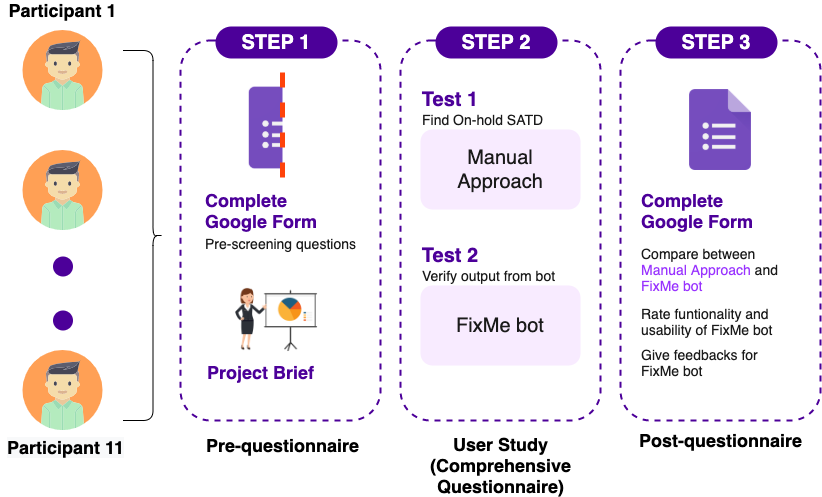}
        \caption{Evaluation setting}
        \label{fig:user_evaluation}
    \end{figure}
    
    \subsection{Evaluation Setup}
    To evaluate the FixMe bot, we focus on the usability of the tool. Figure \ref{fig:user_evaluation} illustrates the evaluation process. We conducted the study with 11 participants. Three participants are experienced developers from software companies who have experience in managing technical debt in their projects, and eight participants are computer science students who regularly use GitHub and have experience in software development projects more than six months. A mock-up Java project implementing TF-IDF searcher and Jaccard searcher is stimulated on GitHub for our user study. There are 8 Java source code files containing both regular code comments and 9 On-hold SATD comments in total (e.g., // TODO: Use this for now then modify this once https://github.com/mockito/mockito/issues/769 is fixed), and 4 unresolved issue reports. The participants need to work on identifying On-hold SATD comments and check the status of the related issue reports to identify Ready-to-be-fixed On-hold SATDs. The participants must perform the task manually for one hour before we introduce the FixMe bot for them to use. They then need to validate the results from FixMe whether they agree with the On-hold SATD comments identified by FixMe. After that, they fill the questionnaire for the study. The evaluation focuses on five criteria using a five point Likert’s scale (from strongly disagree to strongly agree). We would like to evaluate 1) the participants' agreement with the output from FixMe, 2) understandability of the output, 3) ease of use, 4) helping on saving time, and 5) its potential to be adopted in real software projects. In addition, the participants give the overall satisfaction score scaled from 1 to 10 (from very poor to excellent).
    
    
   
\subsection{Result}
We separately report the evaluation results from three professional software developers and eight computer science students as shown in~Table~\ref{tab:result}. Overall, FixMe achieves over 9 out of 10 in the satisfaction score --- 9.2 on average across two~groups. 

\textit{Professional software developers}:
The professional software developers found that the first assigned task in manual identification of Ready-to-be-fixed On-hold SATD code comments are moderately difficult (4/5). After using FixMe, the results indicate that they are satisfied with the supports provided by FixMe with an overall satisfaction score of 9.3. The FixMe bot also receive high scores in all evaluation aspects by achieving over 4 in all evaluation criteria. 
   
\textit{Computer science students}:
The results indicate that the output provided by FixMe in forms of pull request comment, commit comment, and issue notification are understandable (4.7/5). The overall satisfaction score is 9 out of 10.

\section{Discussion}
From the questionnaires, it shows that software developers are aware of addressing SATD but the way to tackle SATD in practice consumes tremendous effort from software developers. All participants indicate that finding On-hold SATD comments without using any supporting tool is a non-trivial job. They must go through all comments and check the related issue report's statuses. From the evaluation results, it can also infer that our FixMe bot can provide advantages in terms of time-saving and ease of use that help them realized Ready-to-be-fixed On-hold SATDs easier and faster. Moreover, the evaluation results show that the tool can potentially be useful and adopted in the real~projects.
    
    

    \begin{table}[]
        \caption{Evaluation results}
        \begin{tabular*}{\linewidth}
        {@{\extracolsep{\fill}}p{0.4\linewidth}>{\centering\arraybackslash}p{1.7cm}>{\centering\arraybackslash}p{1.3cm}c@{}}
        \toprule
        Criteria (1 to 5) & Software Developers & CS Students & AVG. \\
        \midrule
        FixMe provides an accurate result & 4.7 & 4.1 & 4.4 \\
        A result is understandable & 4.4 & 4.7 & 4.6 \\
        FixMe is easy to use & 4.3 & 4.5 & 4.4 \\ 
        Using FixMe is time-saving & 5.0 & 5.0 & 5.0 \\ 
        FixMe can be used in real projects & 4.7 & 4.5 & 4.6 \\\midrule
        Satisfaction score (1 to 10) & 9.3 & 9.0 & 9.2 \\
        \bottomrule
        \end{tabular*}
        \label{tab:result}
    \end{table}

\section{Conclusion}
We introduce the FixMe bot which is an On-hold SATD detection tool seamlessly integrated with GitHub to assist developers in monitoring issue status and notify whenever On-hold SATD becomes ready-to-be-fixed. We adopt the existing state-of-the-art On-hold SATD classification model using N-gram technique and text filtering techniques complemented with automated machine learning approach (Auto-sklearn). The FixMe bot allows developers to perform bot configuration according to their project environment and practice including detection techniques, target branch selection, and notification method selection through the bot's web application. Our tool can notify the developers in forms of a pull request comment, commit comment, and issue creation. We performed a usability testing to evaluate the performance of our tool with 11 participants including experienced software developers and computer science students who actively use GitHub. The evaluation results of the FixMe bot shows promising adoption of the tool. We found that FixMe bot can support developers in managing and detecting On-hold SATD resulting in less effort required. Additionally, the bot helps saving time and gives more information to handle On-hold SATDs in their projects. Our future work involve adding a capability of FixMe to work with other programming languages and also identify issue reports from other issue tracking platforms such as JIRA.

\textbf{Acknowledgement.} We thank anonymous reviewers for their constructive
feedback. This  research  is  partially  supported  by  the  Japan  Societyfor
the  Promotion  of  Science  [Grants-in-Aid  for  ScientificResearch
(S) (No.20H05706)].

\bibliographystyle{IEEEtran}
\bibliography{bibtex}




\end{document}